# Revisiting Nyquist-Like Impedance-Based Criteria for Converter-Based AC Systems

Chongbin Zhao, *Student Member, IEEE*, Qirong Jiang, and Yixin Guo

***Abstract*—Multiple types of Nyquist-like impedance-based criteria are utilized for the small-signal stability analysis of converter-based AC systems. It is usually considered that the determinant-based criterion can determine the overall stability of a system while the eigenvalue-based criterion can give more insights into the mechanism of the instability. This paper specifies such understandings starting with the zero-pole calculation of impedance matrices obtained by state-spaces with the Smith-McMillan form, then clarifying the absolute reliability of determinant-based criterion with the common assumption for impedance-based analysis that each subsystem can stably operate before the interconnection. However, ambiguities do exist for the eigenvalue-based criterion when an anticlockwise encirclement around the origin is observed in the Nyquist plot. To this end, a logarithmic derivative-based criterion to directly identify the system modes using the frequency responses of loop impedances is proposed, which owns a solid theoretical basis of the Schur complement of transfer function matrices. The theoretical analysis is validated using a PSCAD simulation of a grid-connected two-level voltage source converter.**

## I. INTRODUCTION

REGARDING the power system small-signal analysis, the state-space based method is recognized as a standard method [1]. It first requires establishing a set of time-domain nonlinear equations [2], then linearization and autonomy, and finally the eigenvalue calculation of the state matrix [3]-[5]. However, considering the tendency of vendor privacy protection in converter-dominated systems, it becomes less and less feasible to build a high-fidelity state-space model.

To this end, the impedance-based method which is effective and practical in the black-box system, has attracted tremendous research interests in both the academic and the industry. This method allows vendors to stabilize their design under ideal grids and the utilities to examine the grid-converter interactions using frequency responses. Learning from the successful practices in DC systems [6], [7], the impedance-based criterion has been extended to AC systems in the recent ten years, e.g., guiding the wind farm constructions in China [8].

The so-called "impedance/admittance" refers to a frequency-domain transfer function that describes the small-signal ratio between voltage and current, thus the impedance-ratio criterion can be concluded in a Nyquist-like framework based on Cauchy's argument principle for graphical analyses [9]-[11]. Since the Nyquist plots of open-loop systems are used to identify the closed-loop stability, the *frame coupling* [12], [13] of AC rather than DC system, will lead to high-dimensional open-loop transfer function matrices and the use of generalized Nyquist criterion [10]. In addition, the extra requirement of the system autonomy for AC rather than DC system introduces widely-used *dq* and sequence domain impedance models based on the techniques of Park transformation and harmonic balance, respectively [14]-[16].

By aggregating the open-loop grid- and converter-side impedance (matrix) as the return (impedance) differences/ratio [17], two methods are reported for stability analysis [18], [19]. The first method calculates the determinant and counts the number of encirclements of the unique locus around the origin in the Nyquist plot, which is believed to tell whether an unstable mode exists in the system. The second method calculates the eigenvalues, and once a specific locus encircles the origin, the corresponding physical loop should be responsible for the instability. This paper will offer a group of examples to question the universality of the eigenvalue-based criterion while the reliability of the determinant-based criterion is theoretically proved.

Obtaining the *Smith-McMillan form* [18] by diagonalization is a standard method to acquire the zeros, poles, and determinant of a transfer function matrix. Nevertheless, the approach of diagonalizing a transfer function matrix is not unique, e.g., the submatrix with Schur complement, which yields the decoupled *loop impedance* [20] where partial system modes are reflected in the numerator (polynomial) but extra right-half plane (RHP) poles may exist in the denominator (polynomial), which sacrifices the advantage of "no RHP pole inspection" of the determinant-based criterion; the RHP poles counting methods in [21], [22] are complicated without a direct solution on the problem since only RHP zeros are of concern. On the other hand, if the practical frequency responses of loop impedance measured in the sequence domain are available, performing the vector fitting [23] can theoretically identify the zeros and poles, but the irrational form of sequence domain models does not match the rational transfer function identifications of the algorithms, so the identified zeros and poles are untrusted. The criterion

This work was supported in part by the National Natural Science Foundation of China under Grant U22B6008. *(Corresponding author: Chongbin Zhao).*

C. Zhao and Q. Jiang are with the Department of Electrical Engineering, Tsinghua University, Beijing 100084, China (e-mail: zhaocb19@mails.tsinghua.edu.cn; qrjiang@mail.tsinghua.edu.cn); Y. Guo is with the School of Electrical and Electronic Engineering, North China Electric Power University, Baoding 071000, China (e-mail: 2295865911@qq.com)

proposed in [24] assumes a two-order polynomial factored by a pair of conjugate roots and regarding the rest of the transfer function as a whole. By separating the real and imaginary parts of frequency responses, the targeted unstable modes with positive real parts should satisfy a group of conditions, but [25] points out that such a criterion may misjudge a stable mode to be unstable. Hence, an alternative for the eigenvalue-based criterion to achieve the distributed analysis is still required.

This paper aims at revisiting the origin of the impedance-based method, explicating the relative criteria, and making necessary revisions. It has the following innovative points.

1) Revealing the most appealing point of applying the impedance-based method for system-level analyses [26]-[28]. Using the Smith-McMillan form of transfer function matrices, it is proved that the multiplication or sum of two transfer function matrices without RHP poles surely does not introduce an RHP pole, which benefits the determinant-based criterion.

2) Perfecting a systematical method to obtain impedance models in both *dq* and sequence domains, which guarantees the order of denominator equal to the number of system state variables, avoids the ambiguous zero-pole emergence, and facilitates the correct use of Nyquist-like criteria.

3) Discussing the reliability of the widely used Nyquist-like criteria through the comparative study. Specifically, it is recommended to avoid the eigenvalue-based criterion because of the uncertain emergence of the RHP pole.

4) Proposing a novel criterion to reveal the physical insights of an unstable system. The criterion is based on the logarithmic derivative of loop impedances to directly identify the unstable modes, which equally treats each zero and pole. The application in the system-level analysis is envisioned.

The remainder of this paper is organized as follows. Section II clarifies the theoretical foundation of impedance-based stability criteria. Section III introduces the studied system and derives the impedances using the state-space, then conducts the comparative studies to examine the generality of determinant- and eigenvalue-based criterion. A novel criterion to remedy the existing impedance-based criteria is proposed in Section IV. Section V discusses and concludes the work.

## II. Theoretical Foundation

### *A. Impedance Matrix of Open-Loop Systems*

The two-level voltage source converter (TL-VSC) grid-connected system considered in this paper is three-phase symmetrical without zero-sequence differential-mode dynamics. Even if modeling in the *dq* domain [14] adapts to the common *dq* decoupling control and yields an inherent linear time-invariant model, extra rotations to the grid are required for the system-level analysis or the measurement [15], hence the sequence domain models became mainstream in recent years. The open-loop sequence domain model can be transformed from the corresponding *dq* domain model through a combination of rotation and frequency shift [29], so both the grid- and converter-side impedances are 2×2 matrices regardless of the modeling frame.

Based on the classical control theories, a transfer function matrix can be exported from the state space:

$$\begin{cases}\Delta\dot{\boldsymbol{x}} = \boldsymbol{A}\Delta\boldsymbol{x} + \boldsymbol{B}\Delta\boldsymbol{u} \\ \Delta\boldsymbol{y} = \boldsymbol{C}\Delta\boldsymbol{x} + \boldsymbol{D}\Delta\boldsymbol{u}\end{cases} \tag{1}$$

$$\Delta\boldsymbol{y}(s) = \underbrace{[\boldsymbol{C}(s\mathbf{I}-\boldsymbol{A})^{-1}\boldsymbol{B}+\boldsymbol{D}]}_{\text{Transfer Function Matrix}}\Delta\boldsymbol{u}(s) \tag{2}$$

where $\boldsymbol{A}$, $\boldsymbol{B}$, $\boldsymbol{C}$, $\boldsymbol{D}$, and $\mathbf{I}$ are the constant-value state, input, output, feedforward, and identity matrices, respectively. $\boldsymbol{x}$, $\boldsymbol{y}$, and $\boldsymbol{u}$ are the state, input, and output vectors formed by the corresponding variables, respectively. The prefix Δ represents the small signals while *s* is the Laplace operator.

When TL-VSC serves as an open-loop current-source subsystem and accesses a voltage-source subsystem, the admittance should be established for TL-VSC and applied in the impedance-based stability criterion [16]. Such a preference can be explained as the AC current dynamics are the elements of $\Delta\boldsymbol{x}$ and $\Delta\boldsymbol{y}$ while the AC voltage dynamics cannot be included in $\Delta\boldsymbol{y}$, so (2) yields the admittance, and the impedance is the inversion of admittance. In addition, since the eigenvalues of $\boldsymbol{A}$ reflect the system modes, indicating by $(s\mathbf{I}-\boldsymbol{A})^{-1}$, all the elements in the admittance (not the impedance) model hold the same denominator that is factored by the system modes, and a stable open-loop subsystem surely does not induce an RHP pole of each element in the *primary* transfer function matrix.

Even if the above conclusions are often regarded as cognitions for impedance-based analysis, clarifying a basic concept requires a series of theoretical derivations, which should be the "starting point" of revisiting the stability criteria. Moreover, one should attach great importance to the conclusions given by the state-space method.

### *B. Smith-McMillan Form of Transfer Function Matrix*

Considering that the stable operation of each device under ideal grids is a common premise of the impedance-based method, even if Section II. A explains no RHP pole exists in any element of the primary matrix for each open-loop subsystem, whether an RHP pole exists in the complete open-loop system must be proved by the demonstration using Smith-McMillan forms of transfer function matrices, which distinguishes AC and DC systems. Supposing that the grid-/converter-side impedance/admittance $\boldsymbol{Z}_g/\boldsymbol{Y}_c$ own a general form regardless of the modeling frame:

$$\begin{aligned}\boldsymbol{Z}_g(s) &\triangleq \begin{bmatrix} N_g^{11}(s)/D_g(s) & N_g^{12}(s)/D_g(s) \\ N_g^{21}(s)/D_g(s) & N_g^{22}(s)/D_g(s)\end{bmatrix} \triangleq \boldsymbol{N}_g(s)/D_g(s),\ \boldsymbol{Y}_g(s) = \boldsymbol{Z}_g^{-1}(s), \\ \boldsymbol{Y}_c(s) &\triangleq \begin{bmatrix} N_c^{11}(s)/D_c(s) & N_c^{12}(s)/D_c(s) \\ N_c^{21}(s)/D_c(s) & N_c^{22}(s)/D_c(s)\end{bmatrix} \triangleq \boldsymbol{N}_c(s)/D_c(s),\ \boldsymbol{Z}_c(s) = \boldsymbol{Y}_c^{-1}(s).\end{aligned} \tag{3}$$

$$\boldsymbol{R}_{ZY} = \boldsymbol{Z}_g\boldsymbol{Y}_c(or = \boldsymbol{Y}_c\boldsymbol{Z}_g) = \boldsymbol{N}_g\boldsymbol{N}_c/D_gD_c \triangleq \boldsymbol{N}_{gc}/D_{gc},\ \boldsymbol{R}_{ZY}{}' = \boldsymbol{Z}_c\boldsymbol{Y}_g(or = \boldsymbol{Y}_g\boldsymbol{Z}_c). \tag{4}$$

where $\boldsymbol{R}_{ZY}/\boldsymbol{R}_{ZY}{}'$ is the recommended/not recommended return ratio [16] and the symbol "(*s*)" is mostly neglected starting from (4) for simplicity. Obtaining the Smith-McMillan form of $\boldsymbol{R}_{ZY}$, i.e., $\boldsymbol{R}_{ZY}^{SM}$, requires three steps:

- Factoring out the least common multiple of all the denominators of elements in $\boldsymbol{R}_{ZY}$, that is $D_{gc}$.
- Finding the Smith-Normal form of $\boldsymbol{N}_{gc}$ by solving the

greatest common divisor $\chi_i$ ($i$ =0, 1, 2) of all $i\times i$ minor determinants and computing $\epsilon_i=\chi_i/\chi_{i-1}$ ($\chi_0$=1).

- $\boldsymbol{R}_{\rm ZY}^{\rm SM}$ is simplified from diag($\epsilon_1/D_{\rm gc}$, $\epsilon_2/D_{\rm gc}$) to diag($\epsilon_1/\delta_1$, $\epsilon_2/\delta_2$) considering the zero-pole cancellation.

The zeros and poles of $\boldsymbol{R}_{\rm ZY}^{\rm SM}$ are defined as the roots of $\epsilon_1\epsilon_2$ and $\delta_1\delta_2$, respectively. The determinant of ($\mathbf{I}$+$\boldsymbol{R}_{\rm ZY}$), whose numerator can factor out the modes [18], is calculated as:

$$\begin{aligned}&\det(\boldsymbol{R}_{\rm ZY}) = k_1k_2\det(\boldsymbol{R}_{\rm ZY}^{\rm SM}) = k_1\epsilon_1k_2\epsilon_2/\delta_1\delta_2\\&\Rightarrow \det(\mathbf{I}+\boldsymbol{R}_{\rm ZY}) = \det(D_{\rm gc}\mathbf{I}+\boldsymbol{N}_{\rm gc})/D_{\rm gc} = (\epsilon_1+k_1\delta_1)(\epsilon_2+k_2\delta_2)/\delta_1\delta_2\end{aligned} \quad (5)$$

where $k_1$ and $k_2$ are constants. Therefore, it is concluded that $\boldsymbol{R}_{\rm ZY}$, or more importantly, det($\mathbf{I}$+$\boldsymbol{R}_{\rm ZY}$), does not have an RHP pole as long as each subsystem $\boldsymbol{Z}_{\rm g}$ or $\boldsymbol{Y}_{\rm c}$ does not have an RHP pole. Such an idea can be extended to the system-level analysis by partitioning the equivalent grid- and converter-side subsystems, developing the *transfer immittance* matrix of each device, and forming high-dimensional partitioned matrices for stability analysis [26]-[28]. The derivations above using the transfer function matrices instead of their generalized expressions [17], [19], [21] also lay a more rigorous basis.

The determinants of return differences ($\boldsymbol{S}_{\rm Z}$ (=$\boldsymbol{Z}_{\rm g}$+$\boldsymbol{Z}_{\rm c}$) or $\boldsymbol{S}_{\rm Y}$ (=$\boldsymbol{Y}_{\rm g}$+$\boldsymbol{Y}_{\rm c}$)) and ($\mathbf{I}$+$\boldsymbol{R}_{\rm ZY}'$) can also be deduced:

$$\begin{aligned}&\det(\boldsymbol{S}_{\rm Z})=\det(D_{\rm gc}\mathbf{I}+\boldsymbol{N}_{\rm gc})/[D_{\rm g}\det(\boldsymbol{N}_{\rm c})],\\&\det(\boldsymbol{S}_{\rm Y})=\det(D_{\rm gc}\mathbf{I}+\boldsymbol{N}_{\rm gc})/[D_{\rm c}\det(\boldsymbol{N}_{\rm g})],\\&\det(\mathbf{I}+\boldsymbol{R}_{\rm ZY}')=\det(D_{\rm gc}\mathbf{I}+\boldsymbol{N}_{\rm gc})/[\det(\boldsymbol{N}_{\rm gc})].\end{aligned} \quad (6)$$

Based on (5) and (6), the four return differences and ratios have the same set of zeros which can reflect the closed-loop stability, but only det($\mathbf{I}$+$\boldsymbol{R}_{\rm ZY}$) definitively does not hold an RHP pole. That is to say, only det($\mathbf{I}$+$\boldsymbol{R}_{\rm ZY}$) is reliable for identifying the overall stability with the unique characteristic loci in the Nyquist plot, which helps avoid solving the characteristic equation (det($\mathbf{I}$+$\boldsymbol{R}_{\rm ZY}$)=0). Moreover, the extra design principle in [17] to ensure det($\boldsymbol{N}_{\rm c}$) without RHP root is redundant and unrealistic for a transfer function matrix.

## *C. Eigenvalue- and Schur Complement-Based Diagonalization*

The process of obtaining the Smith-McMillan form is essentially a diagonalization:

$$\mathbf{I}+\boldsymbol{R}_{\rm ZY}^{\rm SM} = \boldsymbol{L}\times(\mathbf{I}+\boldsymbol{R}_{\rm ZY})\times\boldsymbol{R} \quad (7)$$

where $\boldsymbol{L}$ and $\boldsymbol{R}$ are unimodular matrices that keep the zeros /poles of det($\mathbf{I}$+$\boldsymbol{R}_{\rm ZY}$) equal to that of det($\mathbf{I}$+$\boldsymbol{R}_{\rm ZY}^{\rm SM}$). The following two methods are also essentially diagonalizations. The first method is given by the eigenvalue decomposition of ($\mathbf{I}$+$\boldsymbol{R}_{\rm ZY}$), which yields an explicit equation of $\lambda_{\rm ZY}^{1,2}$ for a 2×2 matrix:

$$\lambda_{\rm ZY}^{1,2} = 1+\frac{N_{\rm gc}^{11}+N_{\rm gc}^{22}\pm\sqrt{(N_{\rm gc}^{11}-N_{\rm gc}^{22})^2+4N_{\rm gc}^{12}N_{\rm gc}^{21}}}{2D_{\rm gc}} \quad (8)$$

In the practical application, only the frequency responses of $\lambda_{\rm ZY}^{1,2}$ are available by substituting the frequencies of interest into ($\mathbf{I}$+$\boldsymbol{R}_{\rm ZY}$) and performing the numerical calculations. Even if one expects to obtain the information of zeros of ($\mathbf{I}$+$\boldsymbol{R}_{\rm ZY}$) using $\lambda_{\rm ZY}^{1,2}$ through Nyquist plots, there is no evidence that a perfect transfer function can be obtained from the square root in (8) [18], which may introduce uncertainty to the stability information and take away the "no RHP condition" of $\lambda_{\rm ZY}^{1,2}$ indicated by $D_{\rm gc}$.

The second method is based on the process of deducing the Schur complement of a submatrix in ($\mathbf{I}$+$\boldsymbol{R}_{\rm ZY}$):

$$\begin{aligned}&\begin{bmatrix}\underbrace{D_{\rm gc}+N_{\rm gc}^{11}-N_{\rm gc}^{12}N_{\rm gc}^{21}/(D_{\rm gc}+N_{\rm gc}^{22})}_{\det(D_{\rm gc}\mathbf{I}+\boldsymbol{N}_{\rm gc})/(D_{\rm gc}+N_{\rm gc}^{22})} & 0\\ 0 & D_{\rm gc}+N_{\rm gc}^{22}\end{bmatrix}/D_{\rm gc}\\&=\begin{bmatrix}1 & -N_{\rm gc}^{12}/(D_{\rm gc}+N_{\rm gc}^{22})\\0 & 1\end{bmatrix}\begin{bmatrix}D_{\rm gc}+N_{\rm gc}^{11} & N_{\rm gc}^{12}\\N_{\rm gc}^{21} & D_{\rm gc}+N_{\rm gc}^{22}\end{bmatrix}\begin{bmatrix}1 & 0\\-N_{\rm gc}^{21}/(D_{\rm gc}+N_{\rm gc}^{22}) & 1\end{bmatrix}/D_{\rm gc}\end{aligned} \quad (9)$$

The derivations of Schur complements for $\boldsymbol{S}_{\rm Z}$ and $\boldsymbol{S}_{\rm Y}$ can be learned from (9) and thus neglected. Taking $\boldsymbol{S}_{\rm Z}$ and frame "1" as an example, the Schur complement (the upper-left element) after transformation is defined as the loop impedance with clear physical meanings revealed by Fig. 1:

$$S_{\rm Z}^1 = -\Delta u^1/\Delta i^1 = Z_{\rm g}^{11}+Z_{\rm c}^{11}-(Z_{\rm g}^{12}+Z_{\rm c}^{12})(Z_{\rm g}^{21}+Z_{\rm c}^{21})/(Z_{\rm g}^{22}+Z_{\rm c}^{22}). \quad (10)$$

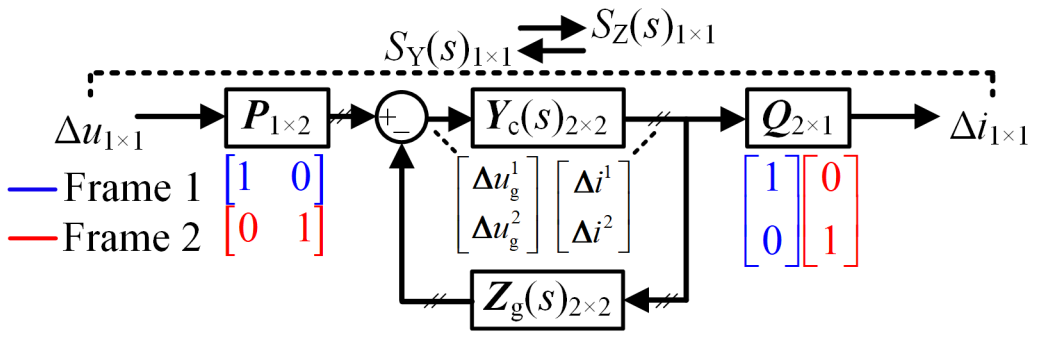

Fig. 1. A generalized representation of the closed-loop system.

A critical observation is that the numerators of all types of Schur complements are equal to those of the determinants in (5) and (6) for the 2×2 matrices, but the feature of RHP poles for each Schur complement is not consistent with that of det($\mathbf{I}$+$\boldsymbol{R}_{\rm ZY}$). Hence, Nyquist-like criteria are not suitable to be applied to the Schur complements. In [20], $S_{\rm Z}^1$ is decomposed to the equivalent grid-/converter-side impedance $Z_{\rm c}$/$Z_{\rm g}$.:

$$\Delta u^1 = \Delta u_{\rm c}^1 - \Delta u_{\rm g}^1 \Rightarrow \begin{cases}Z_{\rm c}^1 = -\Delta u_{\rm c}^1/\Delta i^1 = Z_{\rm c}^{11}-Z_{\rm c}^{12}(Z_{\rm g}^{21}+Z_{\rm c}^{21})/(Z_{\rm g}^{22}+Z_{\rm c}^{22})\\Z_{\rm g}^1 = \Delta u_{\rm g}^1/\Delta i^1 = Z_{\rm g}^{11}-Z_{\rm g}^{12}(Z_{\rm g}^{21}+Z_{\rm c}^{21})/(Z_{\rm g}^{22}+Z_{\rm c}^{22})\end{cases}, \quad (11)$$
$$R_{\rm ZY}^1 = Z_{\rm g}^1/Z_{\rm c}^1,\ S_{\rm Z}^1 = Z_{\rm c}^1+Z_{\rm g}^1.$$

Such an idea is prevailing since it was proposed. In [8], the open-loop impedance matrices are first field-tested and transformed based on (10) and (11) for stability analyses, and the results are used to determine whether the positive- or negative-sequence circuit is responsible for the instability. However, RHP poles of $Z_{\rm c}$ and $Z_{\rm g}$ must be separately inspected before plotting the characteristic loci even if both subsystems can stably operate under ideal grids, and the valuable application of loop impedance requires a deeper investigation.

# III. VALIDATION OF EXISTING IMPEDANCE-BASED CRITERIA

## *A. System Overview*

With the theoretical specification in Section II, the basic TL-VSC connects to an AC weak grid is focused on in this paper. Only the DC-bus voltage control, the AC current control, and the phase-locked loop (PLL) are considered in the grid-following control of Fig. 2, which is simple but suits this work because:

- PLL is a decisive cause of sequence-domain frequency coupling for the converter-side impedance models.
- By counting the integral outputs, AC inductor currents, and DC capacitor voltage, the system order is 8, which is quite low and benefits the zero/pole calculation.
- The obtained AC impedance models are *fully observable* to the complete modes of the closed-loop system and can

analyze arbitrary operating conditions.

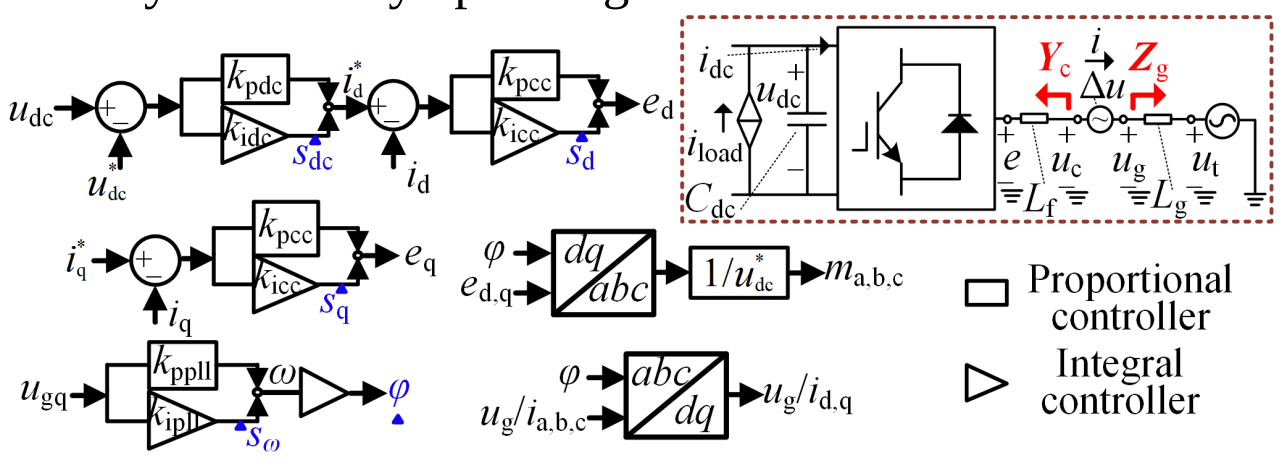

Fig. 2. Control loops. Superscript * represents reference.

TABLE I
KEY PARAMETERS OF SYSTEM

| Symbols | Case A | Case B |
|---|---|---|
| $u_t$, $u_{dc}^*$, $\omega_0$ | 380∠0° V, 750 V, 2π50 rad/s | |
| $L_f$, $L_g$, $C_{dc}$ | 0.55 mH, 0.8/0.9/1.0 mH, 5 mF | 5 mH, 2.1/2.2/2.3 mH, 5 mF |
| $i_{load}$ | 66.66 A | 3.1 A |
| $i_q^*$ | 0 A | 5 A |
| $k_{pdc}+k_{idc}/s$ | 0.1+100/$s$ | 1+100/$s$ |
| $k_{pcc}+k_{icc}/s$ | 0.1+10/$s$ | 1+1000/$s$ |
| $k_{ppll}+k_{ipll}/s$ | 3+100/$s$ | 5+2000/$s$ |

By connecting the harmonic signal graphs of the typical nonlinear time-periodic system [32], [33], the conclusions of a TL-VSC can be reasonably extended to other types of converters, such as the modular multilevel converter. In addition, adopting the average model instead of the switching model in the simulation excludes the influence of PWM on stability analyses. Table I lists the key parameters for comparative studies. One can combine Fig. 2 and Table I to understand the symbol meanings.

## *B. Impedance Modeling Using State-Space*

Explicit expressions of impedance models are available using the frequency-domain harmonic linearization for TL-VSCs [30], [31]. However, the inversion of a transfer function matrix often leads to dimension explosion. Hence, the actual order of an impedance model is mostly not equal to its theoretical order, which means some zero-pole pairs can be canceled, or the zeros-poles cannot be factored out. Here a systematic method for obtaining the open-loop impedance or loop impedance using the state space is offered. The complete process is neglected but some notes are added for obtaining high-fidelity models.

- Most state spaces are established in the *dq* domain for a TL-VSC, which is also the starting point of this work. Considering that the partial state variables appear in pairs (e.g. [$s_d$, $s_q$] in Fig. 1) are aligned to the *control dq* frame while the input and output of the impedance model are aligned to the *grid dq* frame, the frame rotation should be included in the impedance modeling but the system modes will not be affected by the rotation.
- The rigorous equilibrium calculation by numerical iteration is the basis of linearization for any device under closed-loop control [25]. Note that there are minute changes in the state equations regarding the AC voltage dynamics between state spaces for obtaining the open-loop impedance and loop impedance, the equilibria should be kept the same based on the closed-loop system.
- The derivation of sequence domain impedance models is recommended to integrate the frame transformation from a *dq* domain real vector to a sequence domain complex vector into ***A***, ***B***, ***C***, and ***D*** [5]:

$$\boldsymbol{A}_{\rm pn}=\boldsymbol{T}_{\bf x}^{-1}(\boldsymbol{A}_{\rm dq}+{\rm j}\omega_0\boldsymbol{I})\boldsymbol{T}_{\bf x},\ \boldsymbol{B}_{\rm pn}=\boldsymbol{T}_{\bf x}^{-1}\boldsymbol{B}_{\rm dq}\boldsymbol{T}_{\bf u},\ \boldsymbol{C}_{\rm pn}=\boldsymbol{T}_{\bf y}^{-1}\boldsymbol{C}_{\rm dq}\boldsymbol{T}_{\bf x},\ \boldsymbol{D}_{\rm pn}=\boldsymbol{T}_{\bf y}^{-1}\boldsymbol{D}_{\rm dq}\boldsymbol{T}_{\bf u}. \tag{12}$$

where the subscript dq/pn represents the *dq*/sequence domain. ***T*** represents the rotation, whose order is indicated by the subscript vectors and owns several submatrices 0.5×[1, 1; –j, j] regarding the *dq* pairs. j is the imaginary unit, $\omega_0$ is the rated angular frequency, and j$\omega_0$**I** represents the frequency shift.

Note that even if multiplying 0.5×[1, 1; –j, j] with the *dq* domain impedance model and substituting *s* to ($s$–j$\omega_0$) can also yield the sequence domain impedance model [29], the obtained transfer functions also suffer the dimension explosion; using MATLAB command *mineral* to obtain the authentic zeros and poles requires trial and error. The proposed derivation fully avoids such an issue except in the case of (11), since the one-dimensional equivalent grid-/converter-side impedance model is a virtual concept and does not correspond to an actual system as well as a specific state-space.

## *C. Pole Check of Open-Loop Gain*

Fig. 3 presents the distributions of system modes for both cases in Table I when the TL-VSC accesses to ideal AC grid, where the grid inductor variation leads to the various operating conditions and modes. No negative damping mode is observed, i.e., no RHP pole exists for the primary output transfer functions matrices of the open-loop gain, which indicates that both cases can be used for the stability analysis of subsystem interconnections.

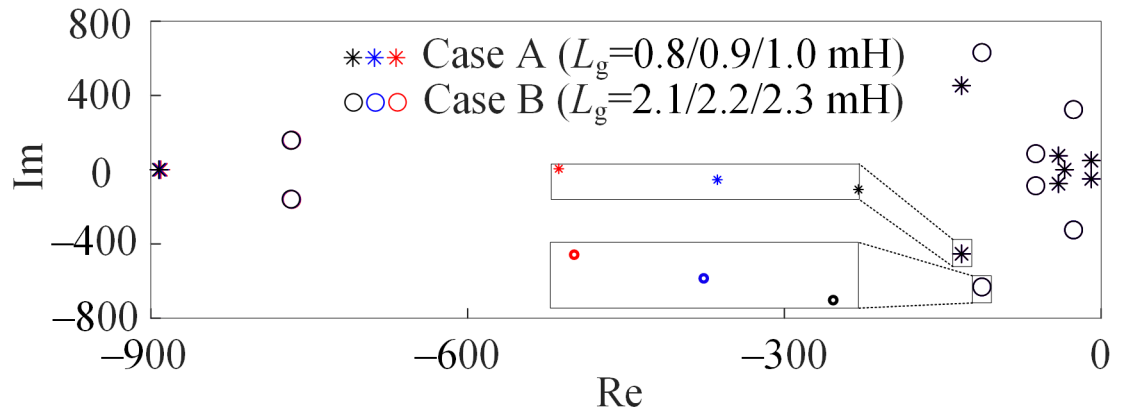

Fig. 3. Distributions of system modes (*dq* domain). The corresponding modes are very close and almost overlap.

The time-domain simulation performed in PSCAD as Fig. 4 shows, where only the waveforms of $u_{dc}$ are presented for their representativeness. The closed-loop system becomes unstable only if $L_g$ steps to the maximum setting values for both cases. Such phenomena conform to the cognition that the system tends to be unstable due to the multiple interaction patterns between the PLL and weak grid [34].

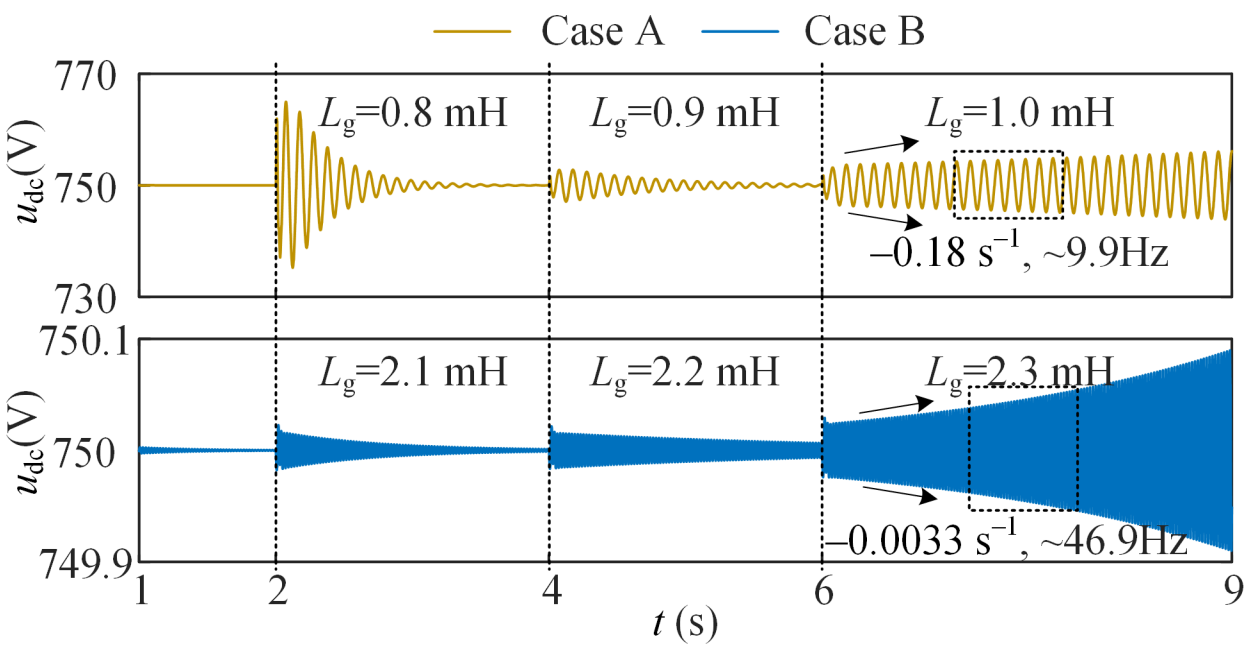

Fig. 4. Simulation waveforms of subsystem interconnections.

### *D. Comparative Study on Impedance-based Criteria for Distributed Stability Analysis*

This subsection will reveal the ambiguity of impedance-based criteria for distributed stability analysis. Both sequence and *dq* domain impedance models are applied with discernible differences found for the two studied cases.

Fig. 5 illustrates the Nyquist plots of Case A. Thereinafter "1-d" represents the characteristic locus of (1+$R_{ZY}$) given by (11), while "2-d" represents the characteristic loci of ($\mathbf{I}$+$\boldsymbol{R}_{ZY}$) given by (4). Because of the mutual conjugate symmetry about j$\omega_0$ for sequence impedances and the self-symmetry about 0 for *dq* impedances, only a single locus is shown for each sequence domain plot from –50 to 150 Hz, while two loci distinguished by the solid and dotted line are shown for each *dq* domain plot from –100 to 100 Hz. In addition, regarding the number of encirclements around (0, j0), it is different for each locus in 1-d plots, but the total encirclement should be counted for 2-d plots especially doubling the number for the sequence domain plots [18].

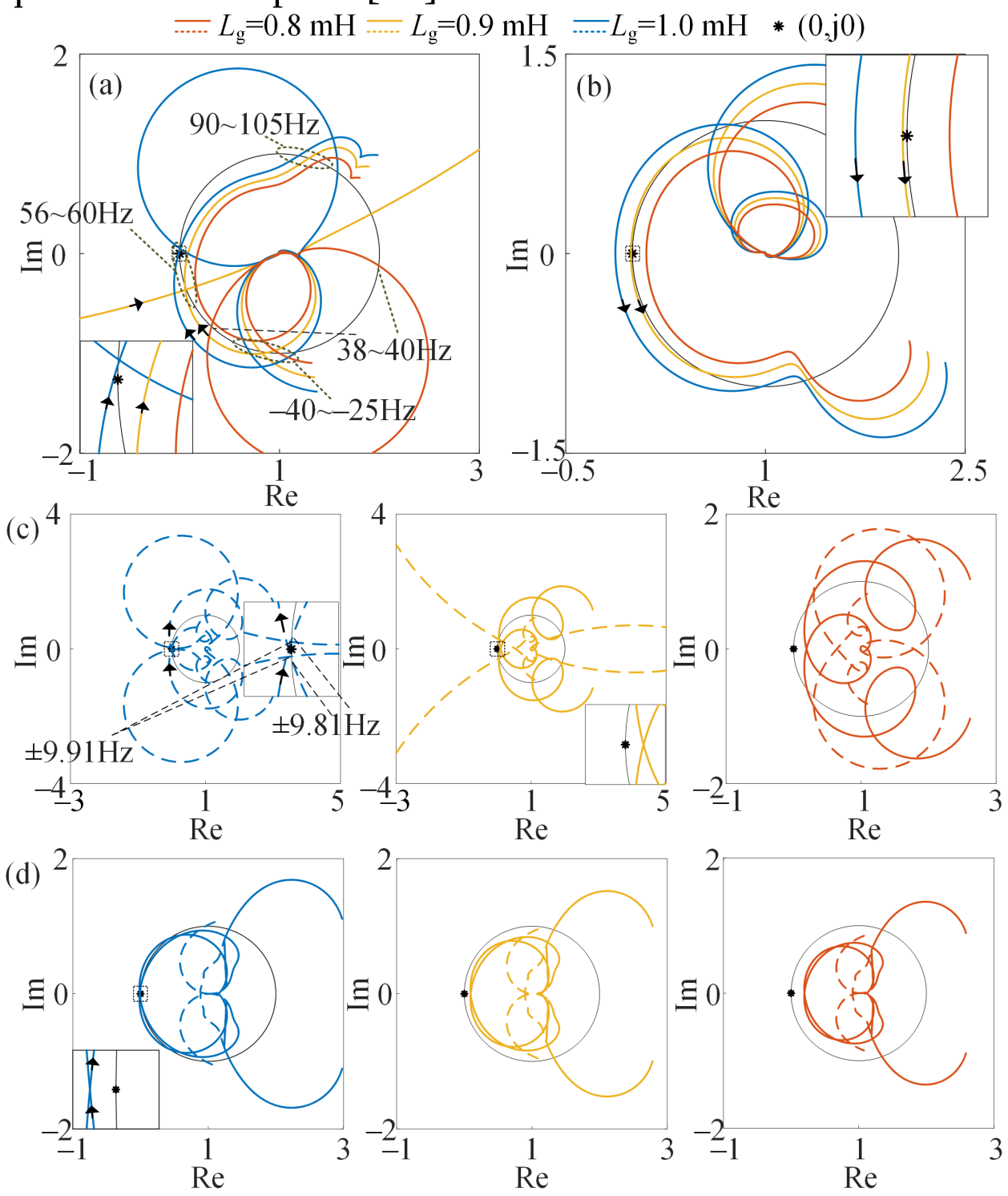

Fig. 5. Nyquist plots for Case A. (a)-(b) 1-d and 2-d in sequence domain. (c)-(d)1-d and 2-d in *dq* domain.

Recalling that only setting $L_g$ =1.0 mH can lead the system to be unstable in Fig. 4, the particularity of Case A is discussed below:

- Regarding the sequence domain 1-d plot in Fig. 5 (a), there is one clockwise encirclement when $L_g$=1.0 mH but one anticlockwise encirclement when $L_g$=0.9 mH. Whereas it can be concluded that the system is unstable when $L_g$=1.0 mH, it cannot be concluded that the system is stable at $L_g$=0.9 mH other than by inspecting the RHP poles of $R_{ZY}$ (equals to that of 1+$R_{ZY}$), which is reflected by Table II. It is observed that one RHP pole exists when $L_g$=1.0 and 0.9 mH while two RHP zeros exist when $L_g$=1.0 mH, which confirms the uncertainty of open-loop RHP pole induced by the Schur complement-based matrix diagonalization and explains the difference between locus. Because of the conjugate symmetry, there is no reason to determine whether the positive or negative sequence loop induces instability from the Nyquist plots.
- Regarding the sequence domain 2-d plot in Fig. 5 (b), it is very confusing to use it for stability analysis because the feature of two anticlockwise encirclements does not change when $L_g$ decreases from 1.0 mH to 0.9 mH, and the instability when $L_g$=1.0 mH cannot be ascertained. See Section IV. D for a deeper discussion on the observation.
- Regarding the *dq* domain analysis, Fig. 5 (c) shows that for the 1-d Nyquist plot, when $L_g$=1.0 mH, there are two clockwise encirclements around the origin for the *d*-locus (the solid line) instead of the *q*-locus (the dotted line). One may infer that the *d*-frame control leads to the instability, but it is not rigorous because Table II shows that (1+$R^q_{ZY}$) has two zeros and two poles, and no single encirclement just manifests the system to be unstable, which indicates the importance of determining the RHP poles when using the Schur complement-based criterion. While for the 2-d Nyquist plots in Fig. 5 (d), the change of loci with the variation of $L_g$ distinguishes from Fig. 5 (b), which is common in existing graphical analyses.

TABLE II
Zeros and Poles OF 1+$R_{ZY}$ (×10$^2$, Case A)

| | | | | | | | |
|---|---|---|---|---|---|---|---|
| $L_g$=1.0 mH | *Z* | *d, q* | **0.0018±j0.62** | –0.61±j4.07 | –0.27±j0.48 | –0.35 | –2.26 |
| | *P* | *d* | –1.02±j4.72 | –0.38±j0.93 | –0.10±j0.49 | –0.35 | –2.19 |
| | | *q* | –0.94±j3.77 | –0.30±j0.44 | **0.10±j0.58** | –0.35 | –8.91 |
| | *Z* | *p, n* | **0.0018+j3.76** | –0.61+j7.21 | –0.27+j3.62 | –0.35+j3.14 | –2.26+j3.14 |
| | | | **0.0018+j2.52** | –0.61–j0.93 | –0.27+j2.67 | | |
| | *P* | *p* | –5.29+j3.12 | –0.55+j7.18 | –1.72–j1.02 | –0.41+j3.76 | |
| | | | –0.068+j3.63 | –0.35+j3.13 | –0.28+j2.67 | **0.024+j2.48** | |
| | | *n* | –5.29+j3.17 | –0.55–j0.90 | –1.72+j7.31 | –0.41+j2.52 | |
| | | | –0.068+j2.65 | –0.35+j3.15 | –0.28+j3.61 | **0.024+j3.80** | |
| $L_g$=0.9 mH | *Z* | *d, q* | –0.015±j0.62 | –0.66±j4.10 | –0.27±j0.49 | –0.35 | –2.46 |
| | *P* | *d* | –1.05±j4.73 | –0.40±j0.91 | –0.10±j0.49 | –0.35 | –2.41 |
| | | *q* | –0.95±j3.81 | –0.30±j0.45 | 0.002±j0.58 | –0.35 | –8.92 |
| | *Z* | *p, n* | –0.015+j3.76 | –0.66–j0.96 | –0.27+j3.63 | –0.35+j3.14 | –2.46+j3.14 |
| | | | –0.015+j2.52 | –0.66+j7.24 | –0.27+j2.65 | | |
| | *P* | *p* | –5.43+j3.10 | –0.58+j7.22 | –1.69–j1.03 | –0.41+j3.77 | |
| | | | –0.068+j3.63 | –0.35+j3.13 | –0.27+j2.66 | **0.0020+j2.48** | |
| | | *n* | –5.43+j3.19 | –0.58–j0.94 | –1.69+j7.32 | –0.41+j2.51 | |
| | | | –0.068+j2.65 | –0.35+j3.15 | –0.27+j3.62 | **0.0020+j3.80** | |

Note: *Z*/*P* represents the zero/pole while *d*/*p* and *q*/*n* correspond to frames 1 and 2 in the *dq*/sequence domain models, respectively. There is no RHP zero or pole when $L_g$=0.8 mH. The RHP zeros and poles are in bold. The imaginary parts are consistent (*dq* domain) or complementary about 100π (sequence domain) for the zeros in the same row.

- The preference of selecting a certain intersection point between the loci and unit cycle as the unstable mode may not work for Case A, especially for the sequence domain plot in Fig. 5 (a), because none of the frequencies of intersection points correspond to the simulated (50±9.9) Hz of AC waveforms as Fig. 4 indicates, and the obtained phase margins for the same operating condition between various criteria are not consistent. Thus, it may be questionable for the quantitative information like phase margin or participation factor [35], [36] given by the existing Nyquist-like criteria apart from judging whether the system is stable based on classical control theories.

To test whether the above findings are general, Case B is well-studied with a similar process to Case A. On the whole, it is intuitive to use each criterion to identify the stability based on Fig. 6 even without the simpler zero-pole distribution of Table III, except the 2-d plot in Fig. 6 (b) since the anticlockwise encirclement around the origin for the 1-d plots are observed, but the feature of anticlockwise encirclement changes when $L_g$ decreases from 2.3 to 2.2 mH, which differs from that in Fig. 5 (b). To sum up, even if Case B matches the existing usage of impedance-based criteria such as in [20], it cannot cover the special Case A, and only by combining the RHP pole check with the encirclement direction can ensure the creditable distributed stability analyses, which deviates from the origin of impedance-based analysis.

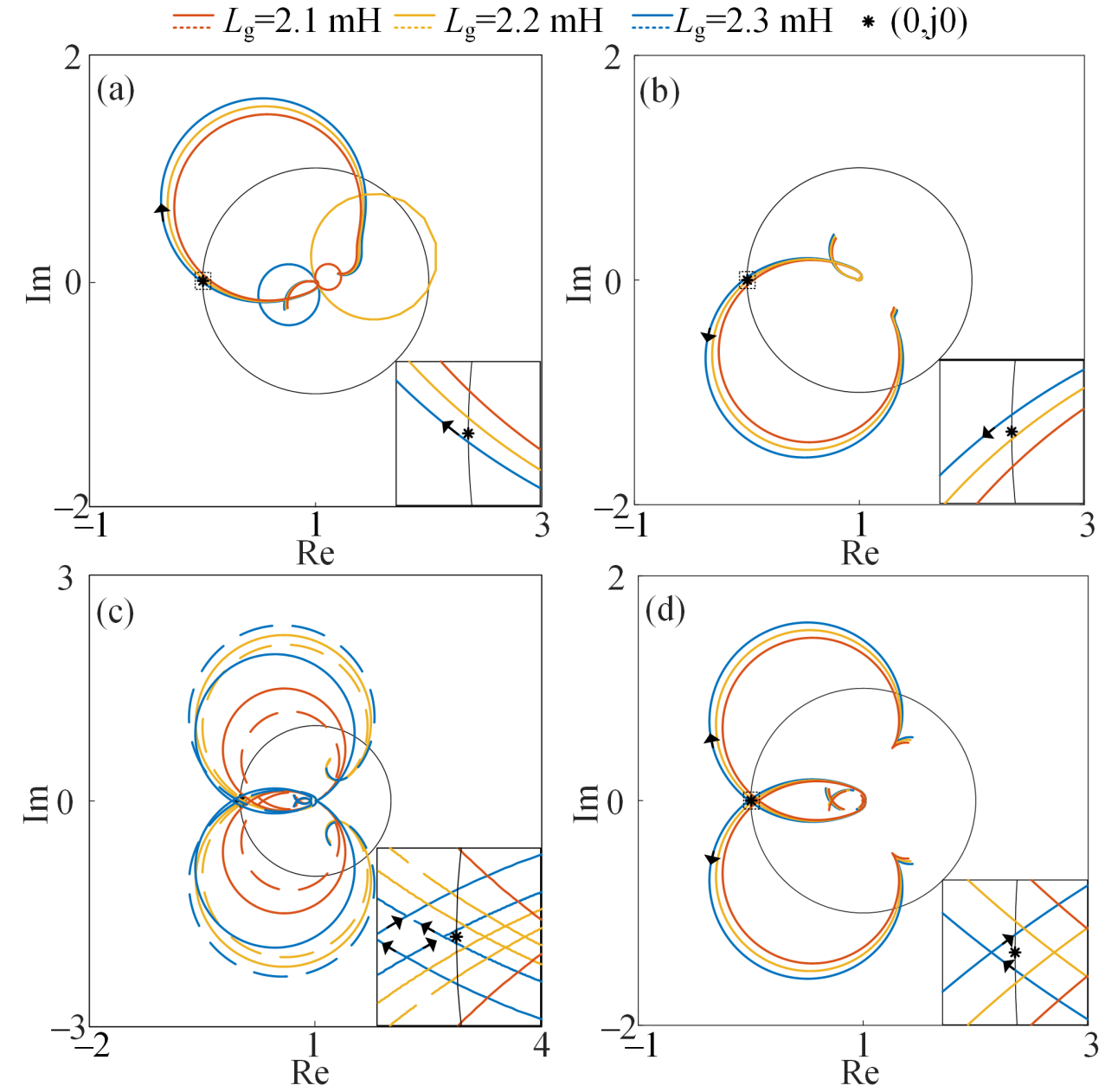

Fig. 6. Nyquist plots for Case B. (a)-(b) 1-d and 2-d in sequence domain. (c)-(d)1-d and 2-d in *dq* domain.

TABLE III

Zeros and Poles OF $1+R_{ZY}$ ($\times 10^2$, Case B)

| | | | | | | |
|---|---|---|---|---|---|---|
| | *Z* | *d*, *q* | **0.0033±j2.94** | –0.63±j0.84 | –0.79±j6.18 | –5.24±j3.35 |
| | *P* | *d* | –0.90±j6.48 | –5.23±j3.35 | –0.22±j3.35 | –0.62±j0.86 |
| | | *q* | –7.67±j1.60 | –1.04±j5.96 | –0.030±j2.89 | –0.63±j0.84 |
| $L_g$= 2.3 mH | *Z* | *p*, *n* | **0.0033+j6.08** | –0.63+j3.98 | –0.79+j9.32 | –5.24+j6.49 |
| | | | **0.0033+j0.20** | –0.63+j2.30 | –0.79–j3.03 | –5.24–j0.20 |
| | *P* | *p* | –0.80+j9.15 | –6.46+j6.04 | –6.36+j0.10 | –1.21–j3.17 |
| | | | –0.26+j6.39 | –0.60+j4.01 | –0.65+j2.31 | **0.0065+j0.20** |
| | | *n* | –0.80–j2.87 | –6.46+j0.24 | –6.36+j6.18 | –1.21–j9.45 |
| | | | –0.26–j0.10 | –0.60+j2.28 | –0.65+j3.98 | **0.0065+j6.08** |

Note: There is no RHP zero or pole when $L_g$=2.1 & 2.2 mH.

### *E. Comparative Study on Impedance-Based Criteria for Overall Stability Analysis*

The only reliable determinant-based criterion using the 2×2 impedance matrix for the overall stability analysis is verified using Fig. 7. Considering the frame transformation, there is only a j$\omega_0$ shift between det($\mathbf{I}+\boldsymbol{R}_{ZY}^{dq}$) and det($\mathbf{I}+\boldsymbol{R}_{ZY}^{pn}$), hence the former is plotted in Fig. 7. There are two clockwise encirclements around the origin when $L_g$ is set as the maximum for both cases, hence the theoretical analysis using Fig. 7 matches simulations in Fig. 4. One should also note the locus is very concise with a stable open-loop system, which benefits the graphical analysis.

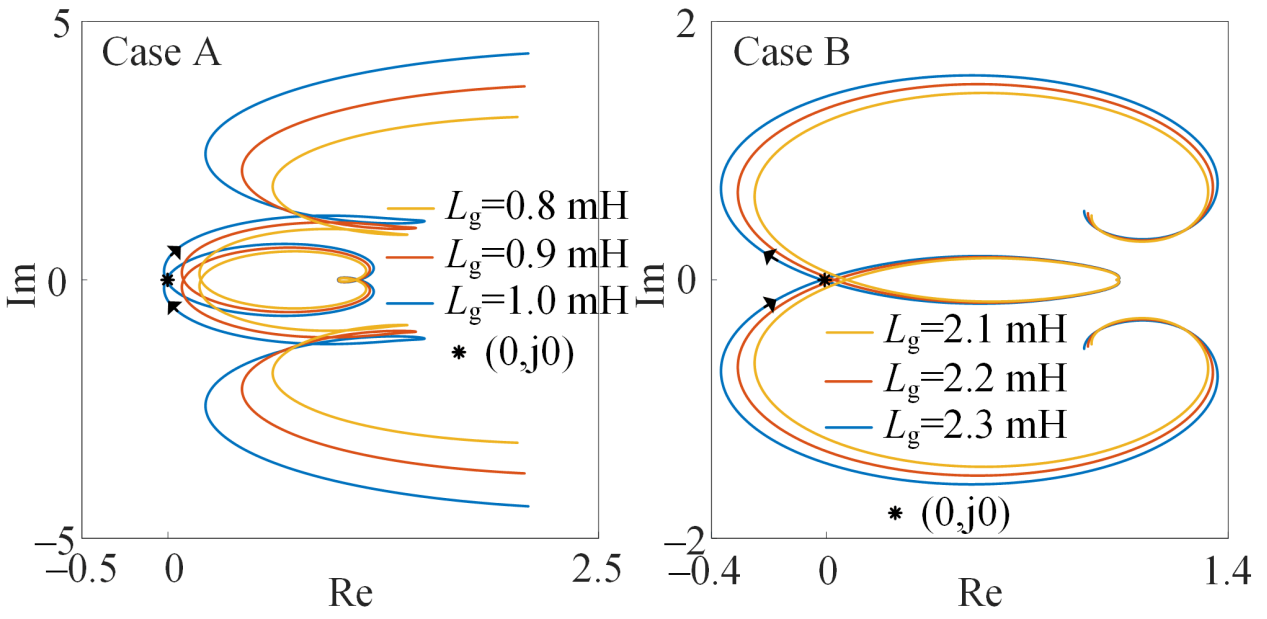

Fig. 7. Stability analysis using det(I+$R_{ZY}^{dq}$).

## IV. Logarithmic Derivative-Based Stability Criterion

### *A. Motivation*

Since there is only one active AC loop in the studied system of this work, the conclusion of overall and distributed stability analyses should be consistent. However, the distributed stability analysis is more of practical interest for positioning a critical loop and guiding a proper oscillation mitigation design in hybrid AC-DC systems, such as the back-to-back system including two AC and one DC subsystems [25]. In [25], the overall instability can be identified by no more than two distributed stability analyses, hence tuning the control parameters which are closely related to the loop that can identify the stability should be the priority. In addition, there are some special cases where the aforementioned open-loop “no RHP pole” condition cannot be ensured, e.g., using a shunt compensation device to stabilize an oscillating system, which even brings challenges to the reliable overall analysis.

Considering that the numerator of the Schur complement can truthfully reflect partial system modes for a high-order system while the uncertainty of positive real part roots for the denominator should be excluded, a novel criterion based on the system mode identification is proposed.

### *B. Basic Principle*

Taking $S_Z$ as an example. Substituting $s$=j$\omega$ into $S_Z(s)$, $S_Z$ can be factored into the zero-pole form:

$$S_Z(\mathrm{j}\omega)=\prod_{i=1}^{n_Z}\mathrm{a}_{Zi}(\mathrm{j}\omega-Z_i)\Big/\prod_{i=1}^{n_P}\mathrm{a}_{Pi}(\mathrm{j}\omega-P_i) \qquad (13)$$

where a, n, $Z$, and P (as well as the subscript) represent the flat gain, the order of polynomials, zeros, and poles, respectively.

Since the essential scope of stability analysis is to identify whether a positive real part $Z_i$ exists in $S_Z$, the basic unit for the numerator, i.e., a first-order polynomial $g_Z(\omega)=a_Z(\mathrm{j}\omega-\lambda_Z)$, $\lambda_Z=\alpha_Z+\mathrm{j}\omega_Z$, is focused on. Compared with the criterion in [24] that requires complex square root operations and irrational approximations, the *logarithmic derivative* on $g_Z$ can eliminate the effect of $a_Z$ and equalize each basic unit:

$$\begin{aligned} D_L(g_Z) &= d\log(g_Z)/d\omega = d(g_Z)/(g_Z d\omega) \\ &= \mathrm{j}/(\mathrm{j}\omega-\lambda_Z) = \mathrm{j}/[-\alpha_Z+\mathrm{j}(\omega-\omega_Z)] \end{aligned} \tag{14}$$

The real-imaginary separation projects a complex output of $D_L(g_Z)$ to two real outputs for quantitative analyses:

$$\begin{cases} \mathrm{Re}[D_L(g_Z)] = (\omega-\omega_Z)/[(\omega-\omega_Z)^2+\alpha_Z^{\,2}] \\ \mathrm{Im}[D_L(g_Z)] = -\alpha_Z/[(\omega-\omega_Z)^2+\alpha_Z^{\,2}] \end{cases}$$
$$\Rightarrow \mathrm{Re}[D_L(g_Z)]\big|_{\omega=\omega_Z}=0,\ \ \mathrm{Im}[D_L(g_Z)]\big|_{\omega=\omega_Z}=-1/\alpha_Z. \tag{15}$$

Eq. (15) shows that a zero-crossing point exists for $\mathrm{Re}[D_L(g_Z)]$ at $\omega=\omega_Z$. To gain insights on $\omega=\omega_Z$, the first and second derivatives of $\mathrm{Re}[D_L(g_Z)]$ and $\mathrm{Im}[D_L(g_Z)]$ are deduced:

$$\begin{cases} \dfrac{d\{\mathrm{Re}[D_L(g_Z)]\}}{d\omega} = \dfrac{-(\omega-\omega_Z)^2+\alpha_Z^{\,2}}{[(\omega-\omega_Z)^2+\alpha_Z^{\,2}]^2} \\ \dfrac{d\{\mathrm{Im}[D_L(g_Z)]\}}{d\omega} = \dfrac{-2(\omega-\omega_Z)^2\alpha_Z}{[(\omega-\omega_Z)^2+\alpha_Z^{\,2}]^2} \end{cases}$$
$$\Rightarrow d\{\mathrm{Re}[D_L(g_Z)]\}/d\omega\big|_{\omega=\omega_Z}=1/\alpha_Z^{\,2},\ d\{\mathrm{Im}[D_L(g_Z)]\}/d\omega\big|_{\omega=\omega_Z}=0,$$
$$\begin{cases} \dfrac{d^2\{\mathrm{Re}[D_L(g_Z)]\}}{d\omega^2} = \dfrac{-2(\omega-\omega_Z)[3\alpha_i^{\,2}-(\omega-\omega_Z)^2]}{[(\omega-\omega_Z)^2+\alpha_Z^{\,2}]^3} \\ \dfrac{d^2\{\mathrm{Im}[D_L(g_Z)]\}}{d\omega^2} = \dfrac{-6(\omega-\omega_Z)^2\alpha_Z^{\,2}+2\alpha_Z^{\,3}}{[(\omega-\omega_Z)^2+\alpha_Z^{\,2}]^3} \end{cases} \tag{16}$$
$$\Rightarrow d^2\{\mathrm{Re}[D_L(g_Z)]\}/d\omega^2\big|_{\omega=\omega_Z}=0,\ d^2\{\mathrm{Im}[D_L(g_Z)]\}/d\omega^2\big|_{\omega=\omega_Z}=2/\alpha_Z^{\,3}.$$

Therefore, when $\alpha_Z>0$, a negative minimum of $\mathrm{Im}[D_L(g_Z)]$ and a positive slope of $\mathrm{Re}[D_L(g_Z)]$ coexist at $\omega=\omega_Z$, and duality is satisfied for $\mathrm{Re}[D_L(1/g_Z)]$ and $\mathrm{Im}[D_L(1/g_Z)]$, where $(1/g_Z)$ can be deemed as a basic unit of the denominator in $S_Z$. The *frequency selectivity* induced by the inverse square term of $\omega$ in (15) suppresses the interactions between zeros and poles, thus the weak negative damping mode can be identified quite exactly using (15) or (16). By substituting $S_Z$ into $D_L(\cdot)$, the process of applying the novel stability criterion is:

1) Obtaining the frequency responses of a dedicated $S_Z$, by either adopting the frequency scans or performing the theoretical deductions based on (9).
2) Calculating the logarithmic derivative of frequency responses using the *difference method* and the first line in (14) with a small enough step such as 0.01Hz.
3) The system is determined as unstable when a minimum of $\mathrm{Im}[D_L(S_Z)]$ and a positive slope of $\mathrm{Re}[D_L(S_Z)]$ coexist at the same frequency ($\omega=\omega_Z$), and $\alpha_Z$ is calculated using the output of $\mathrm{Im}\{D_L[S_Z(\omega_Z)]\}$ and (15).

Since $S_Z$ is a general transfer function that contains the stability information of a closed-loop system, basic ideas of the proposed criterion can be extended to any other transfer-function-based analysis as long as one can find a proper transfer function, such as the loop gain which uses the feedback and reference as output and input, respectively (e.g., $u_{dc}(s)/u^*_{dc}(s)$). More importantly, the proposed criterion can be applied without the RHP pole inspection, but plenty of transfer functions should be selected to ensure the observability of the complete closed-loop system modes, and it is reasonable to select based on the number of physical loops since voltage and current are basic control objectives of power conversion.

### C. Validation Using Loop Impedances

The stability assessments of Case A are illustrated in Fig. 8 for example. It can be observed in Fig. 8 (a), the magnitude of each $S_Z$ owns a pair of nadirs that are symmetrical about the central frequency (0 (50) Hz in *dq* (sequence) domain, and the peaks/nadirs at ±50 Hz in Fig. 8 (b) are due to the calculation errors of the difference method and thus neglected). Consequently, it is more credible that using the various $S_Z$s of the same AC loop but in various domains can achieve the same stability identification rather than the inconsistent feature of loci using unreliable impedance-based criteria shown in Figs. 5 and 6.

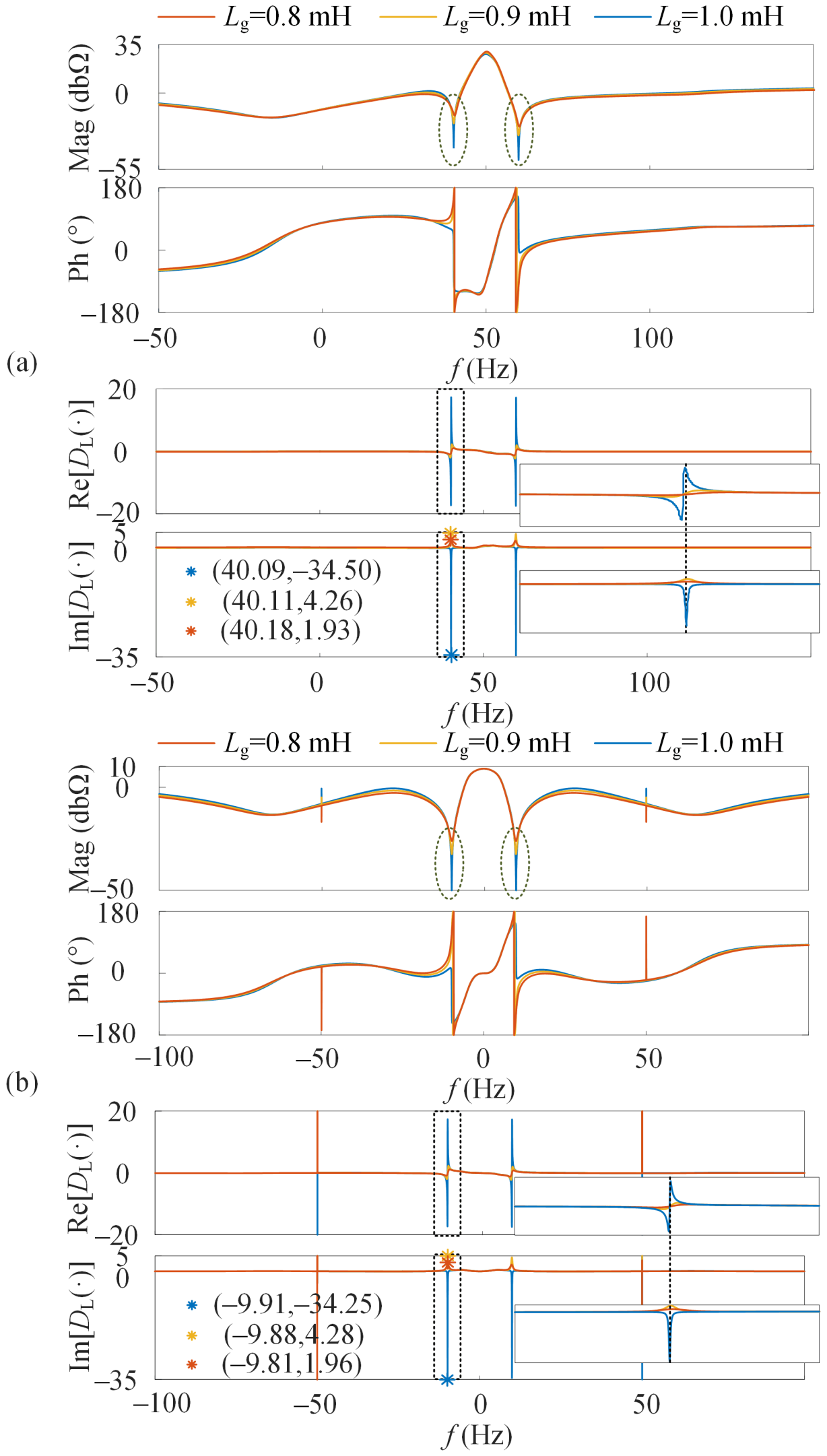

Fig. 8. Impedance characteristics of $S_Z$ with real-imaginary part separations of $D_L(S_Z)$ (Case A). (a) *p*. (b) *d*.

The logarithmic derivatives and real-imaginary part separations are also performed on each $S_Z$ in Fig. 8. When $L_g$ varies, the aforementioned nadirs lead to the zero-crossing points of Re[$D_L(S_Z)$] with definite positive slopes, which means a zero of each $S_Z$ exists over a certain frequency range. Minimums for the case of $L_g$=1 mH and maximums for the case of $L_g$=0.9 & 0.8 mH are observed for Im[$D_L(S_Z)$], and the value of extremes are very close for the models of both domains, thus the system is identified as unstable only when $L_g$=1.0 mH. In addition, $\alpha_Z$ is estimated as 0.1821, –1.4749, and –3.2555 by dividing each extreme value by $2\pi$ (note that the x-axis represents $f$ instead of $\omega$ in Fig. 8) in the sequence domain, and the results match the theoretical calculation in Table II and the simulated divergence rate of $u_{dc}$ in Fig. 4, so the unstable mode can be effectively estimated.

Considering the symmetry of the identified modes about 0 (50) Hz in *dq* (sequence) domain, as Tables II and III indicate, the investigated frequency can be reduced by half for Fig. 8, and only using a single set of plots is enough for each domain, which reduces the computation burden for the proposed criterion. Even if the attribute of frequency selectivity will be influenced when the absolute value of real parts of a zero-pole pair are quite close, using the unconstrained optimization to solve the coefficients of the determined form based on (15) is feasible, and the potential of assessing the system stability over a specific frequency range can be achieved.

### D. Remarks on Eigenvalue-Based Criterion

Considering the symmetry of two sequence domain eigenvalues and the evident discrepancies of two *dq* domain eigenvalues, the ambiguities of the eigenvalue-based criterion in Figs 5 and 6 are revisited by substituting $\lambda_{ZY}^1$ into $D_L(\cdot)$ and performing the real-imaginary part separation in Fig. 9.

It is confirmed that the accurate system modes cannot be reflected by $\lambda_{ZY}$ due to the existence of off-diagonal elements ($N_{gc}^{11}$, $N_{g}^{12}$) and the square root operation in (8). The negative-damping modes are identified as positive-damping modes using the proposed criterion regardless of the encirclement direction in Figs 5 and 6. What's more, there are errors in the stability information, such as the crossover frequency in Fig. 9 (a) of the sequence domain plots, so the quantitative conclusions based on the eigenvalues/eigenvectors [35], [36] are questionable. It is recommended to use the special Case A to study the adaptivity of the generalized Nyquist criterion on an arbitrary converter-based AC system in the field of control theory, but a reliable alternative has been offered in this work.

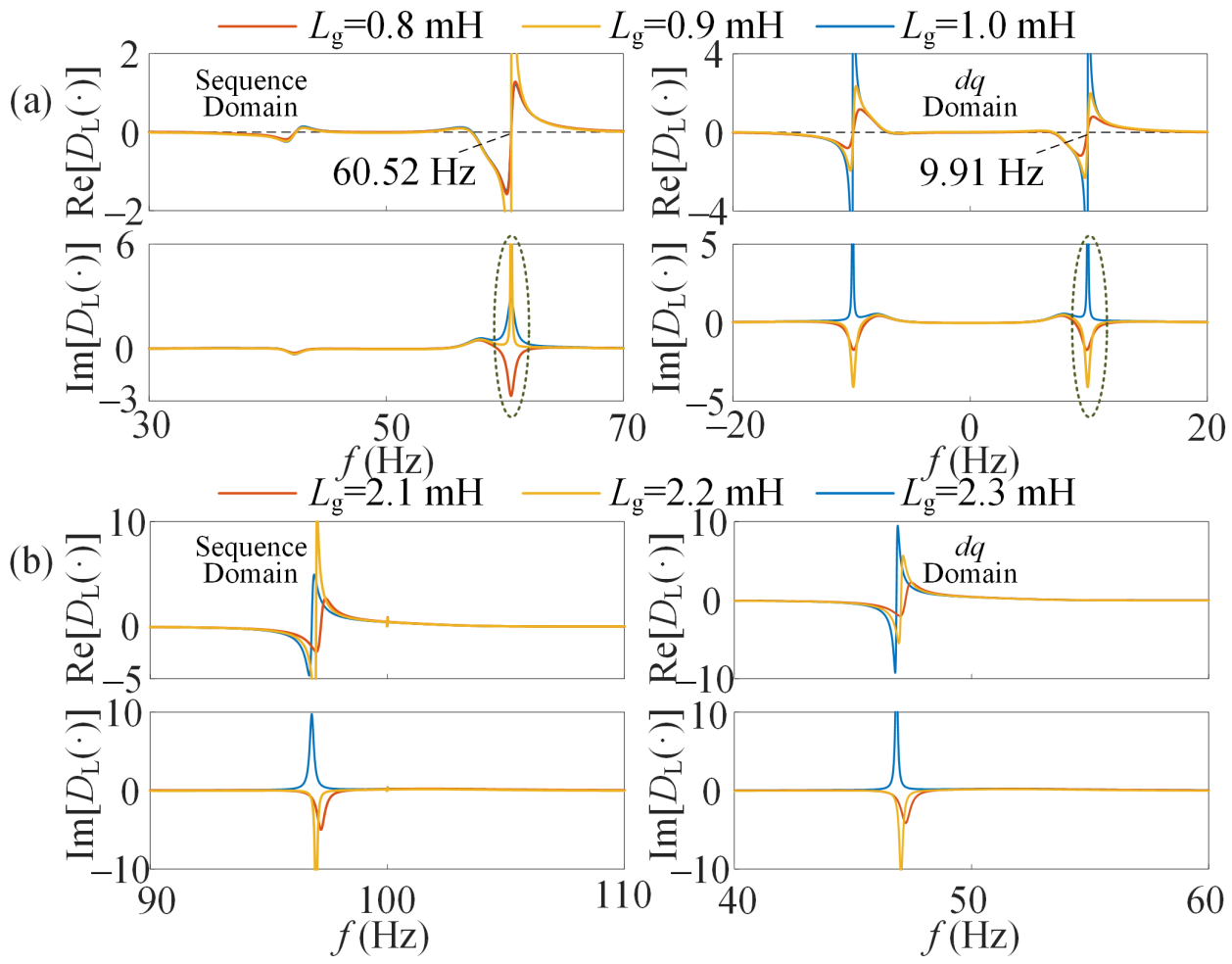

Fig. 9. Real-imaginary part separations of $D_L(\lambda_{ZY}^1)$. (a) Case A. (b) Case B.

## V. Discussion and Conclusion

### A. Overall Stability Analysis Using Transfer Immittances

The concept of transfer immittance is mentioned in Section II. B. Here more comments are offered to help readers correctly regard and apply such a scalable model for system-level analyses. Unlike a 2×2 impedance/admittance matrix where only the dynamics on AC side are selected as the input/output of the open-loop system, the dynamic of DC side (with a superscript 0 below) is added as an element of the input/output vector which yields the 3×3 transfer immittances for a TL-VSC [26]-[28]. [28] states that the input vector of the primary form of transfer immittances should be different when the TL-VSC is under the DC-bus voltage control mode or the (active) power control mode:

$$\text{Power control: } \begin{bmatrix}\Delta i^1\\ \Delta i^2\\ \Delta i^0\end{bmatrix}=\begin{bmatrix}Y^{11} & Y^{12} & Y^{13}\\ Y^{21} & Y^{22} & Y^{23}\\ Y^{31} & Y^{32} & Y^{33}\end{bmatrix}\begin{bmatrix}\Delta v^1\\ \Delta v^2\\ \Delta v^0\end{bmatrix}$$
$$\text{DC-bus voltage control: } \begin{bmatrix}\Delta i^1\\ \Delta i^2\\ \Delta v^0\end{bmatrix}=\begin{bmatrix}Y^{11} & Y^{12} & W^{13}\\ Y^{21} & Y^{22} & W^{23}\\ V^{31} & V^{32} & Z^{33}\end{bmatrix}\begin{bmatrix}\Delta v^1\\ \Delta v^2\\ \Delta i^0\end{bmatrix} \quad (17)$$

It is noticeable that the off-diagonal elements of the third column and row in the transfer immittance for the DC-bus voltage control do not hold the unit of $\Omega^{-1}$. However, [28] does not explain the reason to distinguish the transfer immittances under various control modes even if the conclusion is right. Based on the viewpoint of Section III. B, $\Delta v_0$ is a state variable and also the output variable for the DC-bus voltage control mode, but can only be selected as an input variable for the power control mode. In other words, the two matrices in (17) satisfy the "no RHP pole" condition when a TL-VSC can stably operate fed by ideal sources, and the denominator of each element in (17) is equal to that in (4).

By constructing the diagonal block matrices for both the equivalent converter- and grid-side, the determinant-based impedance criterion can be performed to analyze the overall stability reliably similar to Fig. 8, and one can find an application of such an idea in the Appendix. One can also establish the transfer immittance for the grid-forming converter or the grid-following converter under AC voltage control (the conclusion in [27] is questionable since the AC voltage (amplitude) control should not affect the input/output vectors in (17)), and should always give priority to the physical attribute of voltage source converter instead of a specific control mode for a proper model.

*B. Distributed Stability Analysis by Determinant Decomposing*

The significance of distributed stability analysis is revealed in Section IV. A. Recent work [27] focuses on a 4-terminal DC grid and tries to achieve such a goal by decomposing the system determinant into 5 parts: One part is set to locate the instability at the DC sides with AC side dynamics included, which is similar with the loop impedance; the other four parts locate instability at each AC side but without DC side or the other AC dynamics included but cannot identify the root cause: based on (9), even if the authentic system modes can be reflected in the loop impedance, the lower right element only reflects the attribute of the open-loop system and surely does not reflect any closed-loop system modes; it is inferred that the RHP pole of loop impedance and the RHP zero of the open-loop impedance matrix together result in the questionable root cause identification of [27]. Hence, a mathematical determinant decomposing should not yield a successful distributed analysis theoretically.

Here a modified root cause identification is recommended:

1) Determining the number of physical (AC and DC) loops ($N$) for an interconnected system.
2) Constructing the diagonal block return differences and ratios.
3) Applying the stability criterion of Section IV. B to the $N$ loop using both return differences and ratios for the comparative study, or considering modeling in various domains; once an RHP zero is identified, the physical loop can be regarded as one root cause of instability.

The core idea of the revised identification is the proposed logarithmic derivative-based stability criterion but with more detailed considerations: on one hand, the Schur complement is performed separately for each physical loop, which ensures that the identified RHP zero is the authentic system mode; on the other hand, comparative studies are recommended as Fig. 8 indicates to exclude the *minimal possibility of zero-pole cancellation* due to the potentially emerged RHP pole of loop impedance as (9) and Tables II and III indicates. Counting the loops which identify unstable modes can guide the oscillation mitigation design.

*C. Revisiting Ref. [16]*

It is widely recognized that [16] initiates the impedance-based method in converter-based AC systems over the recent 10 years. [16] majorly clarifies a practical condition of using Nyquist criteria in the power electronics field with the cooperated converter design guideline to ensure such a condition. Since frame coupling is not considered in [16], the obtained criterion for AC systems does not distinguish themselves from those for DC systems [6]. The theoretical foundation in Section II of this work proves that the idea in [16] is valuable and the most appealing point for the practical application of impedance-based methods, especially for the determinant-based criterion for the overall stability analysis.

Unfortunately, lots of so-called "modified impedance-based criteria" (e.g., [20], [27]) obsess with the mathematical equivalence but dismiss the origin of the impedance-based method, especially for the distributed analysis. The particularity of the distributed stability analysis of Case A as shown in Fig. 5 should not be rare based on (8), but it seems that no existing research specifically discusses such an observation since the encirclement direction around the origin is mostly neglected in many Nyquist plots of the literature. Considering that the frequency responses are mostly available, ensuring the open-loop system without an RHP pole is much easier than counting the RHP pole of transfer function functions, and the proposed revised criterion can fully avoid the ambiguity induced by the RHP pole.

*D. Conclusions*

Finally, the contributions of this work are reviewed. It is emphasized that the success of determinant-based overall stability analysis is ensured by the "no RHP condition" of the open-loop system, which is transformed to guarantee the stable operation of the converter- and grid-side subsystems and use the primary form of each transfer function matrix to form impedance ratio. Such a critical condition cannot adapt to the existing impedance-based criteria for distributed stability analyses, which is revealed using both theoretical deductions and Nyquist plots of a special case. A logarithmic derivative-based criterion with only the frequency responses of loop impedance as the input is proposed to serve as a useful tool for the unstable mode identification, which excludes the influence of RHP poles on the graphical stability analyses using Nyquist plots and makes it possible to let the distributed analysis "cover" the overall analysis. It is expected that several findings can guide future device-level modeling and system-level analysis, which are also separately discussed. The applicability of the corresponding conclusions and methods is willing to be examined on higher-order systems using field tests in the future.

## APPENDIX

Fig. 10 reveals a basic topology of point-to-point DC transmission which realizes AC-DC-AC power conversion. If the DC cable is removed, it can be regarded as a back-to-back DC transmission that helps interconnect two AC grids. Here the DC stability is specially focused to emphasize the practical impedance-based criteria, so AC impedances are neglected.

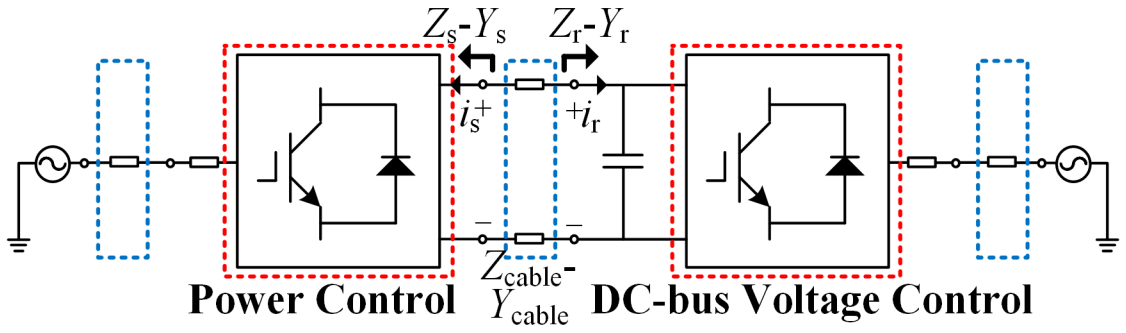

Fig. 10. Topology of point-to-point DC transmission.

For the back-to-back transmission, the well-recognized transfer function for DC stability analysis is (1+$Z_rY_s$), which indicates that if two converters can stably operate under ideal DC-bus voltage (power control) and current (DC voltage control) sources, $Z_rY_s$ does not have an RHP pole and one can draw the Nyquist plots to fast determine the system stability; if using transfer functions such as (1+$Z_sY_r$), ($Z_s$+$Z_r$), or ($Y_s$+$Y_r$), the RHP pole check is unavoidable theoretically before using the corresponding Nyquist plots to correctly identify the system stability. One can be inspired by the following deductions when concluding the above observations into the impedance-based overall and distributed stability analysis framework discussed in the text (neglecting the denominators for simplicity):

$$\begin{bmatrix} v_r \\ i_s \end{bmatrix} = \underbrace{\begin{bmatrix} Z_r & 0 \\ 0 & Y_s \end{bmatrix}}_{\boldsymbol{Z}_{con}} \begin{bmatrix} i_r \\ v_s \end{bmatrix}, \begin{bmatrix} i_r \\ v_s \end{bmatrix} = \underbrace{\begin{bmatrix} 0 & -1 \\ 1 & 0 \end{bmatrix}}_{\boldsymbol{Y}_g} \begin{bmatrix} v_r \\ i_s \end{bmatrix},$$
$$\text{Overall: } \det(1+\boldsymbol{Z}_{con}\boldsymbol{Y}_g) = \det(1+Z_rY_s) = 1+Z_rY_s. \tag{18}$$
$$\text{Distributed: } \boldsymbol{Z}_{con} + \boldsymbol{Y}_g^{-1} = \begin{bmatrix} Z_r & 1 \\ -1 & Y_s \end{bmatrix},$$
$$\det(\boldsymbol{Z}_{con} + \boldsymbol{Y}_g^{-1}) = \det(Z_r + Z_s)\times\det(Y_s) = \det(Y_r + Y_s)\times\det(Z_r).$$

It is intuitive that both the overall and distributed stability analyses can reflect the system mode but cannot embody the advantage of modular analysis using transfer immittances (which are simplified to DC impedance models due to the fix of AC dynamics) over the direct analysis using DC impedance models. However, when the case turns to the point-to-point transmission, the partition point for correctly adopting the DC impedance-based analysis should be calibrated, and a reliable transfer function should be [1+($Z_{cable}$+$Z_r$)$Y_s$], which can be explained using transfer immittance as below:

$$\begin{bmatrix} v_r \\ i_s \end{bmatrix} = \underbrace{\begin{bmatrix} Z_r & 0 \\ 0 & Y_s \end{bmatrix}}_{\boldsymbol{Z}_{con}} \begin{bmatrix} i_r \\ v_s \end{bmatrix}, \begin{bmatrix} i_r \\ v_s \end{bmatrix} = \underbrace{\begin{bmatrix} 0 & -1 \\ 1 & Z_{cable} \end{bmatrix}}_{\boldsymbol{Y}_g} \begin{bmatrix} v_r \\ i_s \end{bmatrix},$$
$$\text{Overall: } \det(1+\boldsymbol{Z}_{con}\boldsymbol{Y}_g) = \det(\begin{bmatrix} 1 & -Z_r \\ Y_s & 1+Y_sZ_{cable} \end{bmatrix}) = 1+Y_s(Z_{cable}+Z_r).$$
$$\text{Distributed: } \boldsymbol{Z}_{con} + \boldsymbol{Y}_g^{-1} = \begin{bmatrix} Z_r + Z_{cable} & 1 \\ -1 & Y_s \end{bmatrix}, \tag{19}$$
$$\det(\boldsymbol{Z}_{con} + \boldsymbol{Y}_g^{-1}) = \det(Z_r + Z_{cable} + Z_s)\times\det(Y_s)$$
$$= \det(Y_s + (Z_r + Z_{cable})^{-1})\times\det(Z_r + Z_{cable})$$

Eq. (19) confirms the reliability of the determinant-based criterion when formulating the transfer immittance to a proper pattern without an RHP pole. From the viewpoint of DC impedance analysis, there is no essential difference between the cases of aggregating the DC cable to either converter (i.e., one cannot say a criterion using ($Z_s$+$Z_r$+$Z_g$) or [$Y_s$+($Z_r$+$Z_g$)$^{-1}$] is wrong), but to assess the number of RHP pole is very important. One can also perform the overall stability analysis using open-loop admittance/impedance:

$$\begin{bmatrix} v_r \\ v_s \end{bmatrix} = \underbrace{\begin{bmatrix} Z_r & 0 \\ 0 & Z_s \end{bmatrix}}_{\boldsymbol{Z}_{con}} \begin{bmatrix} i_r \\ i_s \end{bmatrix}, \begin{bmatrix} i_r \\ i_s \end{bmatrix} = \underbrace{\begin{bmatrix} Y_{cable} & -Y_{cable} \\ -Y_{cable} & Y_{cable} \end{bmatrix}}_{\boldsymbol{Y}_g} \begin{bmatrix} v_r \\ v_s \end{bmatrix},$$
$$\text{Overall: } \det(1+\boldsymbol{Z}_{con}\boldsymbol{Y}_g) = \det(\begin{bmatrix} 1+Z_rY_{cable} & -Z_rY_{cable} \\ -Z_sY_{cable} & 1+Z_sY_{cable} \end{bmatrix}) = 1+Y_{cable}(Z_s+Z_r). \tag{20}$$

In (20), due to the uncertain RHP pole distribution of $Z_s$, the reliability of overall stability analysis without RHP pole check is destroyed, which breaks the major advantage of impedance-based analysis. Such an observation emphasizes the necessity of using transfer immittances for a hybrid AC-DC system or a 100% power electronics-dominated system. An extra notation is that when developing an analytical transfer immittance using harmonic linearization, it may not be straightforward to formulate the input/output vector in (17). The transfer immittance can be reformulated based on the operation of deriving Schur complements as shown in (9).

## Acknowledgment

The authors would like to thank Prof. Jian Sun from Rensselaer Polytechnic Institute for his advice on this work.

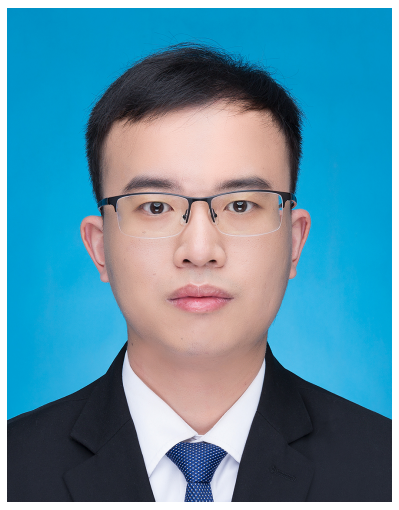

**Chongbin Zhao** (Student Member, IEEE) received the B.S. degree in electrical engineering from Tsinghua University, Beijing, China, in 2019, where he is currently working towards the Ph.D. degree. In 2023, He was a visiting scholar at Rensselaer Polytechnic Institute, Troy, NY, United States. His research interests include power quality analysis and control, and emerging converter-driven power system stability analysis and control.

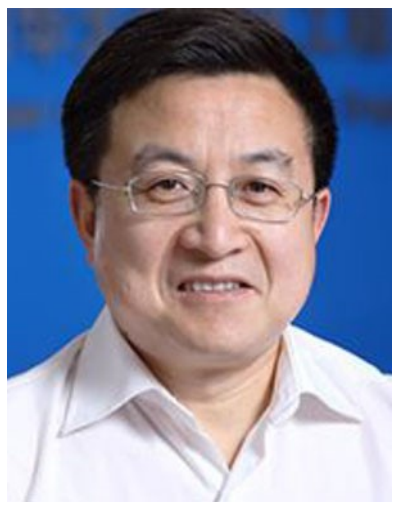

**Qirong Jiang** received the B.S. and Ph.D. degrees in electrical engineering from Tsinghua University, Beijing, China, in 1992 and 1997, respectively. In 1997, he was a Lecturer with the Department of Electrical Engineering, Tsinghua University, where he later became an Associate Professor in 1999. Since 2006, he has been a Professor. His research interests include power system analysis and control, modeling and control of flexible ac transmission systems, power-quality analysis and mitigation, power-electronic equipment, and renewable energy power conversion.

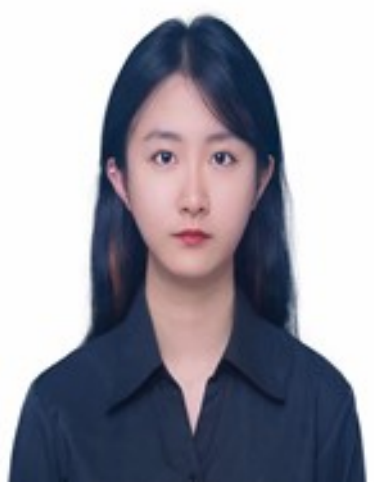

**Yixin Guo** entered the School of Electrical and Electronic Engineering, North China Electric Power University, Baoding, China, in 2020. In 2022, she exchanged in University of Manchester, Manchester, United Kingdom. Her research interests include power system stability analysis.